\documentclass[sigconf,nonacm]{acmart}
\usepackage{float}
\usepackage{graphicx}
\usepackage{caption}
\usepackage{subcaption}
\usepackage{hyperref}

\AtBeginDocument{%
  }

\begin{document}

%%
%% The "title" command has an optional parameter,
%% allowing the author to define a "short title" to be used in page headers.
\title{AI-Enabled Rent-Seeking: How Generative AI Alters Market Transparency and Efficiency}

%%
%% The "author" command and its associated commands are used to define
%% the authors and their affiliations.
%% Of note is the shared affiliation of the first two authors, and the
%% "authornote" and "authornotemark" commands
%% used to denote shared contribution to the research.

\author{YUKUN ZHANG}
\affiliation{%
  \institution{The Chinese University Of Hongkong}
  \city{HongKong}
  \country{China}}
\email{215010026@link.cuhk.edu.cn}

\author{TIANYANG ZHANG}
\affiliation{%
  \institution{University of Bologna}
  \city{Bologna}
 \country{Italy}}
\email{tianyang.zhang@studio.unibo.it}

%%
%% By default, the full list of authors will be used in the page
%% headers. Often, this list is too long, and will overlap
%% other information printed in the page headers. This command allows
%% the author to define a more concise list
%% of authors' names for this purpose.

%%
%% The abstract is a short summary of the work to be presented in the
%% article.
\begin{abstract}
The rapid advancement of generative artificial intelligence (AI) has transformed the information environment, creating both opportunities and challenges. This paper explores how generative AI influences economic rent-seeking behavior and its broader impact on social welfare. We develop a dynamic economic model involving multiple agents who may engage in rent-seeking activities and a regulator aiming to mitigate social welfare losses. Our analysis reveals a dual effect of generative AI: while it reduces traditional information rents by increasing transparency, it also introduces new forms of rent-seeking, such as information manipulation and algorithmic interference. These behaviors can lead to decreased social welfare by exacerbating information asymmetries and misallocating resources. To address these challenges, we propose policy interventions, including taxation and regulatory measures. This study provides a new perspective on the economic implications of generative AI, offering valuable insights for policymakers and laying a foundation for future research on regulating AI-driven economic behaviors.
\end{abstract}

%%
%% The code below is generated by the tool at http://dl.acm.org/ccs.cfm.
%% Please copy and paste the code instead of the example below.
%%

%%
%% Keywords. The author(s) should pick words that accurately describe
%% the work being presented. Separate the keywords with commas.
%\keywords{Generative AI, platform economics, two-sided markets, content creation, winner-takes-all, long-tail effect, game theory, market equilibrium, welfare economics, traffic distribution, social welfare, content platforms, AI-generated content}

%% A "teaser" image appears between the author and affiliation
%% information and the body of the document, and typically spans the
%% page.

%\received{20 February 2007}
%\received[revised]{12 March 2009}
%\received[accepted]{5 June 2009}

%%
%% This command processes the author and affiliation and title
%% information and builds the first part of the formatted document.
\maketitle
\section{Introduction}

\subsection{Research Background and Significance}
The rapid emergence of Generative Artificial Intelligence (generative AI) has fundamentally transformed the digital landscape by enabling systems to autonomously create, manipulate, and disseminate vast quantities of high-quality content. Unlike traditional machine learning models that rely on fixed patterns or predefined outputs, generative AI systems dynamically generate novel content, thereby lowering barriers to information access and democratizing content production. However, this technological revolution also introduces significant risks. As generative AI becomes more pervasive, its potential misuse to distort information ecosystems and facilitate novel forms of rent-seeking behavior is increasingly evident. 

In today’s digital economy, agents can leverage advanced algorithms to exploit information asymmetries, engage in covert content manipulation, and strategically interfere with algorithmic outputs to secure unfair economic advantages. Such practices not only undermine market efficiency and erode public trust but also pose serious threats to social welfare. Recognizing and addressing these emerging challenges is crucial, particularly as AI-driven tools gain prominence in critical areas ranging from financial decision-making to policy formulation.

\subsection{Research Questions and Objectives: The Impact of Generative AI on Rent-Seeking and Social Welfare}
This paper aims to systematically examine how the rise of generative AI reshapes rent-seeking behavior and influences overall social welfare. In particular, we investigate the central question: \textit{Does the proliferation of generative AI give rise to new, information-based rent-seeking strategies, and if so, how do these behaviors affect market efficiency, transparency, and broader social outcomes?}

To address this question, we develop a dynamic economic model that embeds generative AI within the evolving information environment. Our framework considers multiple agents—each with incentives to pursue rent-seeking through both traditional means and innovative, AI-enabled methods. By contrasting the transparency-enhancing effects of generative AI (which can diminish conventional information rents) with its potential to enable more subtle forms of manipulation, our analysis quantifies the net impact on social welfare. This approach not only deepens our understanding of the dual-edged nature of generative AI but also provides a foundation for evaluating policy interventions aimed at mitigating adverse outcomes.

\subsection{Organization of the Paper}
The remainder of this paper is structured as follows. In Section 2, we survey the relevant literature on rent-seeking, information asymmetry, and the emerging influence of generative AI, highlighting gaps that our research seeks to fill. Section 3 introduces our dynamic model, detailing its underlying assumptions, state variables, and functional forms. In Section 4, we conduct a theoretical analysis of the model’s equilibrium outcomes, demonstrating how generative AI modifies rent-seeking incentives and impacts social welfare. Section 5 presents simulation experiments that validate our theoretical insights and explore the effectiveness of various policy interventions.  Finally, Section 7 concludes by summarizing our key findings and outlines directions for future research

\section{Literature Review}
\label{sec:lit_review}

\subsection{Theoretical Foundations of Rent-Seeking Behavior}

Rent-seeking, a concept first introduced by \citet{tullock2008welfare}, refers to efforts by individuals or enterprises to secure economic gains through manipulation of the institutional or informational environment without contributing new value. Early works framed rent-seeking as a source of inefficiency that diverts resources from productive uses, resulting in welfare losses. For example, \citet{acemoglu2015democracy} argue that rent-seeking reinforces class inequality by consolidating wealth among elites, thereby reducing social mobility. Similarly, studies by \citet{rodriguez2004inequality} and \citet{stiglitz2012price} document how rent-seeking exacerbates income inequality through distorted resource allocation.

Despite their contributions, these traditional models exhibit notable limitations. Many assume homogenous agents and static environments, failing to capture the dynamic and heterogeneous nature of modern economies. While \citet{krueger2008political} and \citet{zingales2017towards} emphasize the role of government and institutional structures in shaping rent-seeking, their models often overlook how technological changes can alter rent-seeking incentives. More recent studies, such as those by \citet{catalini2020some} on blockchain and \citet{prat2022attention} on digital markets, have extended the discussion by demonstrating that technological innovations can both reduce traditional information asymmetries and create novel rent-seeking opportunities. However, these works tend to analyze either the efficiency gains or the new distortions in isolation, leaving a gap in our comprehensive understanding of the dynamic interplay between technology and rent-seeking behavior.

\subsection{Development of Generative AI and Its Impact on the Information Environment}

The emergence of generative AI—exemplified by generative adversarial networks \citep{goodfellow2014generative} and large-scale language models \citep{brown2020language}—marks a fundamental shift from static prediction to active information synthesis. This technology not only enhances content creation but also transforms the underlying information environment by increasing data accessibility and transparency. Research in this area has shown promising applications in labor markets \citep{zarifhonarvar2024economics, hui2024short}, market simulation \citep{takahashi2019modeling, xu2019modeling}, and even mechanism design \citep{calvano2020artificial}.

Nevertheless, the literature on generative AI also reveals significant controversies and limitations. While many studies praise generative models for reducing reliance on manual labeling and democratizing content production, concerns remain about ethical risks such as bias, discrimination, and the potential for misuse \citep{weidinger2022taxonomy, vinuesa2020role}. Critically, the capability of generative AI to produce synthetic content creates new avenues for rent-seeking—enabling agents to manipulate information ecosystems through content manipulation and algorithmic interference. Although some researchers acknowledge these risks, a comprehensive framework that captures both the transparency-enhancing benefits and the novel rent-seeking strategies enabled by generative AI is still lacking.

\subsection{Research Gaps}

In summary, current research lacks a unified, dynamic framework that captures the interplay between traditional and AI-enabled rent-seeking. Our study addresses this gap by proposing a dynamic economic model that accounts for the evolving information environment and the strategic adaptations of agents. This approach not only reveals the limitations of existing theories but also offers valuable insights for policymakers balancing innovation with social welfare.

\section{Model Setup}
\label{sec:model_setup}

In this section, we present a dynamic economic model that captures the interplay between agents' rent-seeking behavior and an evolving information environment influenced by generative AI. We begin by outlining the basic framework and defining the key roles, then describe the state variables and the dynamics of the information environment. Next, we incorporate the impact of generative AI through specific parameters, and finally, we discuss the microfoundations of emerging rent-seeking behaviors. Throughout, we highlight the idealized assumptions of our model and discuss potential extensions for future work.

\subsection{Basic Framework and Role Definitions}
We consider a discrete-time dynamic setting with \( t = 0,1,\dots,T \), where multiple economic agents interact within an information environment that evolves over time. The key players in our framework are \emph{agents (rent-seekers)} and a \emph{regulator}. The agents, indexed by \( i \in \{1,\dots,N\} \), each seek to maximize their economic returns and may engage in rent-seeking by exploiting information asymmetries, manipulating content, or interfering with generative AI algorithms. Their individual actions collectively influence the state of the information environment. In contrast, the regulator is a central authority tasked with safeguarding social welfare by mitigating the adverse impacts of rent-seeking; it may deploy policy tools—such as taxation, regulatory constraints, or technological verification measures—to reduce information asymmetries and limit manipulative activities. This framework establishes the roles and interactions that underlie our subsequent analysis.

\subsection{State Variables and the Dynamics of the Information Environment}
To capture the evolution of the information environment, we introduce a scalar state variable $S_t \in [0,1]$, which represents the \textit{transparency level} at time \( t \). Higher values of \( S_t \) indicate greater transparency and reliability, while lower values suggest increased opacity and informational frictions. The dynamics of \( S_t \) are driven by two primary factors: (1) \emph{transparency enhancement}—where generative AI improves information availability, reduces scarcity, and enhances verifiability; and (2) \emph{rent-seeking distortions}—where agents' manipulative activities, such as content manipulation or algorithmic distortion, introduce noise and misinformation, thereby reducing transparency. Formally, we define the state transition as
\begin{align}
    S_{t+1} &= S_t + \gamma - \delta\, \overline{R_t} + \varepsilon_t, \label{eq:state_transition}
\end{align}
where:
\begin{itemize}
    \item \( \gamma \geq 0 \) denotes the baseline improvement in transparency attributable to generative AI, capturing the intrinsic boost from enhanced content verification and information access.       
    \item \( \overline{R_t} = \frac{1}{N} \sum_{i=1}^{N} R_{i,t} \) is the average rent-seeking effort at time \( t \), with \( R_{i,t} \) representing the individual rent-seeking investment.

    \item \( \delta > 0 \) measures the sensitivity of the information environment to rent-seeking; a higher \( \delta \) implies that even modest rent-seeking can substantially degrade transparency.
    \item \( \varepsilon_t \) is a mean-zero random shock capturing unpredictable external influences (e.g., sudden technological or policy changes).
\end{itemize}

\textbf{Note:} Bounding \( S_t \) within \([0,1]\) is an idealization that simplifies our analysis. In practice, transparency may be multi-dimensional and influenced by additional factors; future work could extend the model to incorporate heterogeneous agents or multiple state variables.

\subsection{\texorpdfstring{Incorporating Generative AI: Parameters \(\gamma\), \(\delta\) and \(\Delta B_{\text{AI}}\)}
{Incorporating Generative AI: Parameters gamma, delta and Delta B AI}}

Building on the state dynamics, we now specify how generative AI influences agents' benefit functions. In an AI-rich context, agents' rent-seeking gains are modified by an incremental term \( \Delta B_{\mathrm{AI}}(R_{i,t}, S_t) \). Here, \( \gamma \) captures the intrinsic transparency boost from generative AI—if rent-seeking were absent, a high \( \gamma \) would drive \( S_t \) upward, reflecting a more informed market. Conversely, \( \delta \) quantifies the environment's vulnerability to rent-seeking, so that a larger \( \delta \) indicates that manipulative actions can more readily offset the benefits of generative AI. The term \( \Delta B_{\mathrm{AI}}(R_{i,t}, S_t) \) represents the additional gains (or losses) from rent-seeking in this context, reflecting two opposing effects: on one hand, generative AI increases the overall supply of information, reducing traditional rents; on the other hand, it enables new forms of rent-seeking through sophisticated manipulation—such as fabricating synthetic content or poisoning training data—that can enhance economic gains. Thus, while \( \gamma \) and \( \delta \) govern the evolution of \( S_t \) at the environmental level, \( \Delta B_{\mathrm{AI}} \) adjusts the cost-benefit calculus for individual agents.

\subsection{Microfoundations of Emerging Rent-Seeking Behaviors: Information Manipulation and Algorithmic Interference}

Traditional rent-seeking typically involves overt activities such as lobbying or restricting access to valuable data. However, the advent of generative AI has enabled more covert strategies. Agents can engage in \emph{information manipulation} by leveraging generative AI to produce large volumes of misleading or synthetic content—such as generating fake news or market rumors to sway public opinion, or fabricating counterfeit expert opinions via synthetic academic papers and technical reports—thereby increasing information asymmetry and complicating the task of discerning truth from fabrication. Additionally, agents may directly target AI systems through \emph{algorithmic interference}. This includes both model poisoning, where malicious data is introduced into training sets to skew AI outputs in favor of specific outcomes, and adversarial attacks, in which vulnerabilities in AI models are exploited to generate biased or false outputs that disrupt critical decision-making processes. The combined effect of these actions is a distortion of the state variable \( S_t \), leading to macro-level inefficiencies and reduced transparency, which in turn reinforce the negative externalities of rent-seeking.

In summary, our model captures the dual nature of generative AI: it has the potential to enhance information transparency while simultaneously enabling new forms of rent-seeking that degrade the information environment. In the following sections, we integrate these components into the agents' optimization problems, analyze the resulting equilibrium outcomes, and explore the implications for social welfare and policy interventions.

\section{Model Solution and Theoretical Analysis}

\subsection{Equilibrium Conditions and Steady-State Analysis}
Having established the equilibrium conditions, we now turn to discuss how generative AI alters the incentive structure for rent-seeking. We focus on a Markovian equilibrium in which each agent's rent-seeking choice \( R_{i,t} \) and the system's state \( S_t \) evolve according to rational expectations. Agents take the evolution of \( S_t \) as given and anticipate that other agents follow similar decision rules.

\paragraph{Agent's Optimization Problem:}
Each agent \( i \) chooses \( R_{i,t} \) to maximize expected discounted net benefits:
\begin{equation}\label{eq:agent_problem}
\max_{\{R_{i,t}\}} \mathbb{E}\left[\sum_{t=0}^{T} \delta^{t} \left(B_{\text{AI}}(R_{i,t}, S_t) - C(R_{i,t})\right)\right],
\end{equation}
where \( B_{\text{AI}}(\cdot) \) incorporates both traditional and AI-influenced rent-seeking benefits, and \( C(R_{i,t}) \) is the cost function. A possible specification for \( B_{\text{AI}} \) and \( C \) is:
\begin{equation}\label{eq:benefit_cost_example}
\begin{aligned}
    B_{\text{AI}}(R_{i,t}, S_t) &= \beta R_{i,t}(1 - S_t) + \Delta B_{\mathrm{AI}}(R_{i,t}, S_t), \\
    C(R_{i,t}) &= c R_{i,t}^2.
\end{aligned}
\end{equation}

Under regularity conditions (e.g., concavity and differentiability), each agent's optimal choice in period \( t \) is characterized by the first-order condition (FOC):
\begin{equation}\label{eq:foc_agent}
\frac{\partial B_{\text{AI}}(R_{i,t}, S_t)}{\partial R_{i,t}} - \frac{\partial C(R_{i,t})}{\partial R_{i,t}} = 0.
\end{equation}

In a symmetric equilibrium, we assume that all agents face identical parameters and choose the same rent-seeking effort, denoted as \( R_{i,t}^* = R_t^* \). Aggregating across all agents, we obtain:
\begin{equation}\label{eq:symmetry_condition}
R_t^* = \frac{1}{N}\sum_{i=1}^{N} R_{i,t}^*.
\end{equation}

\paragraph{Steady-State Equilibrium:}
A steady-state equilibrium \( (R^*, S^*) \) is defined by the conditions:
\begin{align}
R^* &= \arg\max_{R} \left[B_{\text{AI}}(R, S^*) - C(R)\right], \label{eq:steady_state_R}\\
S^* &= S^* + \gamma - \delta R^* + \mathbb{E}[\varepsilon], \label{eq:steady_state_S}
\end{align}
with \( \mathbb{E}[\varepsilon]=0 \). From \eqref{eq:steady_state_S}, we obtain:
\begin{equation}\label{eq:steady_state_solution}
\gamma - \delta R^* = 0 \implies R^* = \frac{\gamma}{\delta}.
\end{equation}

This results in a direct relationship between \( \gamma \) and \( \delta \) in the steady state. By substituting \( R^* \) into the agent's FOC and the definition of \( B_{\text{AI}} \), we can solve for \( S^* \) numerically. The existence and uniqueness of \( S^* \) depend on the parameters and the functional form of \( \Delta B_{\text{AI}} \).

\subsection{Impact Mechanisms of Generative AI on Traditional and Novel Rent-Seeking Incentives}
Generative AI significantly alters the rent-seeking landscape through two primary channels:

\paragraph{Diminishing Traditional Rent-Seeking:}
As generative AI increases information availability, the marginal benefit of controlling scarce data diminishes. In traditional settings, \( B(R, S) \) would be high if \( (1 - S) \) is large. However, with generative AI, \( S_t \) tends to increase, reducing \( (1 - S_t) \) and thus the marginal returns to conventional rent-seeking activities such as withholding critical information.

\sloppy
\paragraph{Enabling Novel Rent-Seeking Behaviors:}
Conversely, generative AI opens avenues for new forms of rent-seeking, captured by \( \Delta B_{\mathrm{AI}}(R_{i,t}, S_t) \). Agents exploit advanced tools to produce false signals, manipulate public opinion, or bias model outputs. This can create a scenario where the direct negative effect of AI on traditional rent-seeking returns is offset—or even surpassed—by the emergence of new, more subtle manipulation-based rents. 

In equilibrium, agents weigh these opposing forces. If generative AI predominantly enhances transparency (i.e., \( \gamma \) is large relative to the complexity of new manipulation strategies), then steady-state rent-seeking may decrease. However, if new manipulation techniques provide lucrative returns, overall rent-seeking \( R^* \) could remain substantial, mitigating the gain from increased transparency and keeping \( S^* \) lower than expected from transparency gains alone.

\subsection{Social Welfare Analysis: Quantifying Resource Misallocation and Opacity}
We now evaluate social welfare implications by examining how rent-seeking and information opacity affect resource allocation and market efficiency. Define the social welfare loss function:
\begin{equation}\label{eq:social_loss_function}
L(R_t, S_t) = \phi R_t^2 + \psi (1 - S_t)^2,
\end{equation}
where \( \phi > 0 \) captures the deadweight loss from resource misallocation due to non-productive rent-seeking, and \( \psi > 0 \) measures the social cost of reduced transparency and informational asymmetries.

\paragraph{Welfare Decomposition:}
- \textit{Resource Misallocation (\( \phi R_t^2 \)):} As rent-seeking diverts capital and effort away from productive activities, it reduces aggregate output. The quadratic form ensures that excessive rent-seeking is increasingly costly.
- \textit{Opacity Costs (\( \psi (1 - S_t)^2 \)):} Lower transparency (\( 1 - S_t \)) complicates price discovery, information-based decision-making, and trust. The social cost rises rapidly as the environment becomes more opaque.

\paragraph{Equilibrium Welfare Evaluation:}
At steady state, substituting \( R^* = \gamma/\delta \) and the corresponding equilibrium \( S^* \), we get:
\begin{equation}\label{eq:equilibrium_loss}
L(R^*, S^*) = \phi \left(\frac{\gamma}{\delta}\right)^2 + \psi (1 - S^*)^2.
\end{equation}

A higher \( \gamma \) (enhanced transparency from generative AI) tends to reduce \( (1 - S^*) \), decreasing the opacity cost. However, if novel rent-seeking avenues are sufficiently profitable, \( R^* \) remains high, mitigating the gain from increased transparency and keeping \( L(R^*, S^*) \) elevated.

\subsection{Policy Tools Analysis: Deriving Optimal Strategies for Taxation and Regulation}
The regulator’s goal is to minimize cumulative discounted social loss:
\begin{equation}\label{eq:regulator_problem}
\min_{\{\tau_t\}} \mathbb{E}\left[\sum_{t=0}^{T} \delta^t L(R_t, S_t)\right].
\end{equation}

We consider two main classes of policy interventions:

\paragraph{Taxation (\( \tau_t \)):}
By imposing a per-unit tax on rent-seeking investments, the regulator effectively increases the marginal cost of rent-seeking:
\begin{equation}\label{eq:tax_effect_foc}
\frac{\partial B_{\text{AI}}}{\partial R_{i,t}} - \frac{\partial C}{\partial R_{i,t}} - \tau_t = 0.
\end{equation}
As \( \tau_t \) increases, agents reduce \( R_{i,t} \) to maintain optimality. The optimal tax \( \tau_t^* \) must balance the reduction in misallocation and opacity against potential side effects, such as discouraging certain informational enhancements or beneficial uses of generative AI.

\paragraph{Regulatory Measures:}
Regulation can impose direct constraints on manipulative actions or algorithmic interference. For instance, restricting certain content generation practices reduces the space for \( \Delta B_{\text{AI}} \) to be positive from malicious rent-seeking. In equilibrium, a well-designed regulation that limits the feasible set of \( R_{i,t} \) could yield:
\begin{equation}\label{eq:regulation_condition}
\begin{split}
R_{i,t} &\in \mathcal{F}(S_t), \text{where}\ \mathcal{F}(S_t) \\
 & \text{excludes highly manipulative strategies.}
\end{split}
\end{equation}

This might be conceptualized as raising an effective barrier that reduces \( \delta \), making transparency less susceptible to rent-seeking or directly capping rent-seeking returns. An optimal regulatory policy would set these constraints to achieve a socially desirable \( (R^*, S^*) \) that minimizes \( L(R^*, S^*) \).

\subsection{Summary of Theoretical Insights}  
Our theoretical analysis shows that the steady-state equilibrium rent-seeking level \( R^* \) depends on the interplay between transparency gains (\( \gamma \)) and rent-sensitivity (\( \delta \)). While generative AI can reduce traditional rent-seeking returns, it may also enable more complex manipulation-based rents. Social welfare hinges on balancing resource misallocation and opacity; although AI improves transparency on average, unchecked novel rent-seeking methods can erode welfare gains. Policy instruments like taxation and regulation are essential for realigning private incentives with social welfare goals. These insights provide the foundation for the subsequent simulation-based analysis, where we will numerically verify the results, explore parameter sensitivities, and evaluate the effectiveness of policy tools.

\section{Experiment and Results}

The experimental framework is built on the background of “Generative AI + Rent-Seeking Behavior” and focuses on the scenario of fake information and intelligence manipulation on social media platforms. This framework captures the dual impact of generative AI in both content generation and moderation, while reflecting the dynamic interplay among information asymmetry, agent rent-seeking, and social welfare losses.

\subsection{Experiment Design}

This experiment simulates a scenario of fake information and intelligence manipulation on social media. Ordinary users browse and publish content for reliable information or entertainment, while rent-seeking agents use generative AI to mass-produce content for economic or political gain. The platform or regulator monitors and governs the content.

To better reflect real-world conditions, the simulation incorporates long-term agent benefits, reputation penalties, and potential "collateral damage" from strong supervision. Multiple rounds and scenarios are used to explore the optimal mix and limits of various policy tools (e.g., taxation, detection, and disclosure).

Different groups are formed based on the use of long-term discounting, reputation mechanisms, and supervision. Key output indicators include the average amount of fake information released, platform transparency, agent reputation distribution, social welfare loss, and false injury rate.The details can be found in the appendix.

\subsection{Analysis of Experimental Results}

From the figures above, it is evident that the system exhibits a characteristic “rapid convergence” pattern during the simulation:

\begin{figure}[ht]
    \centering
    \begin{subfigure}[b]{\columnwidth}
        \centering
        \includegraphics[width=\textwidth]{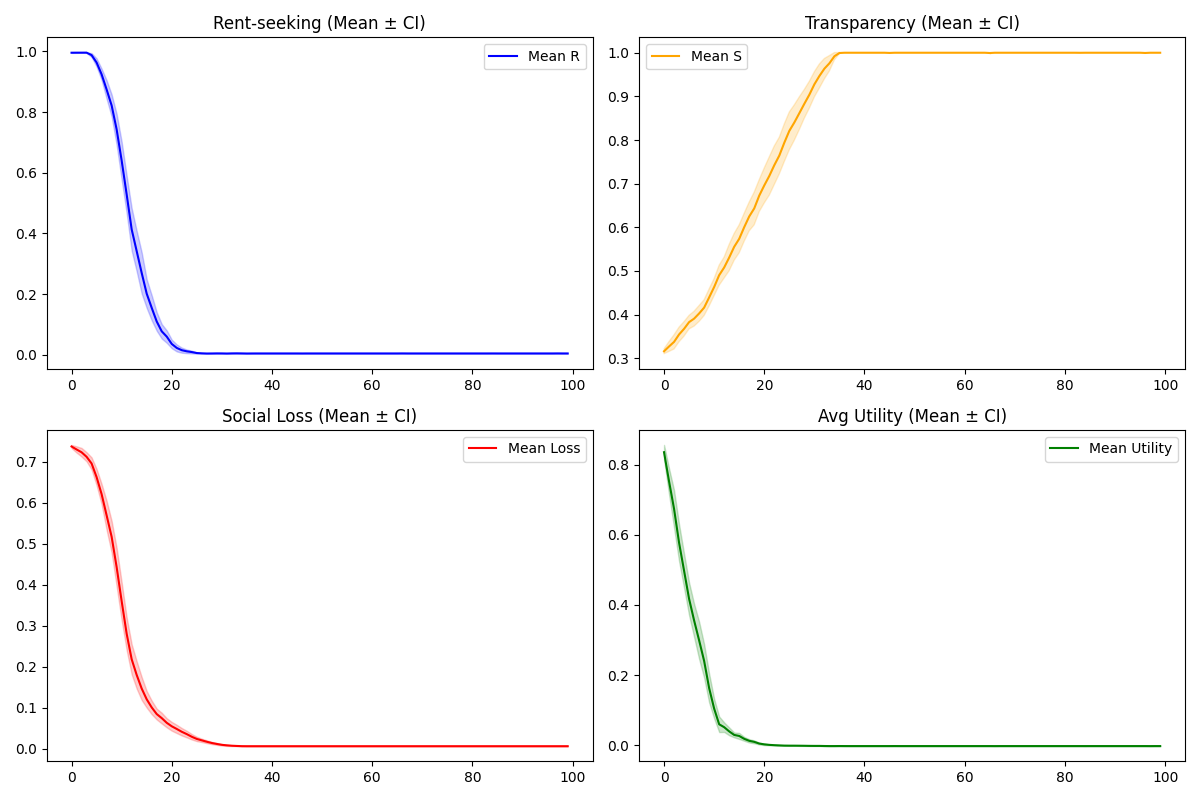}
        \caption{General Experimental Results Part 1.}
        \label{fig:General Experimental Results Part 1}
    \end{subfigure}

    \begin{subfigure}[b]{\columnwidth}
        \centering
        \includegraphics[width=\textwidth]{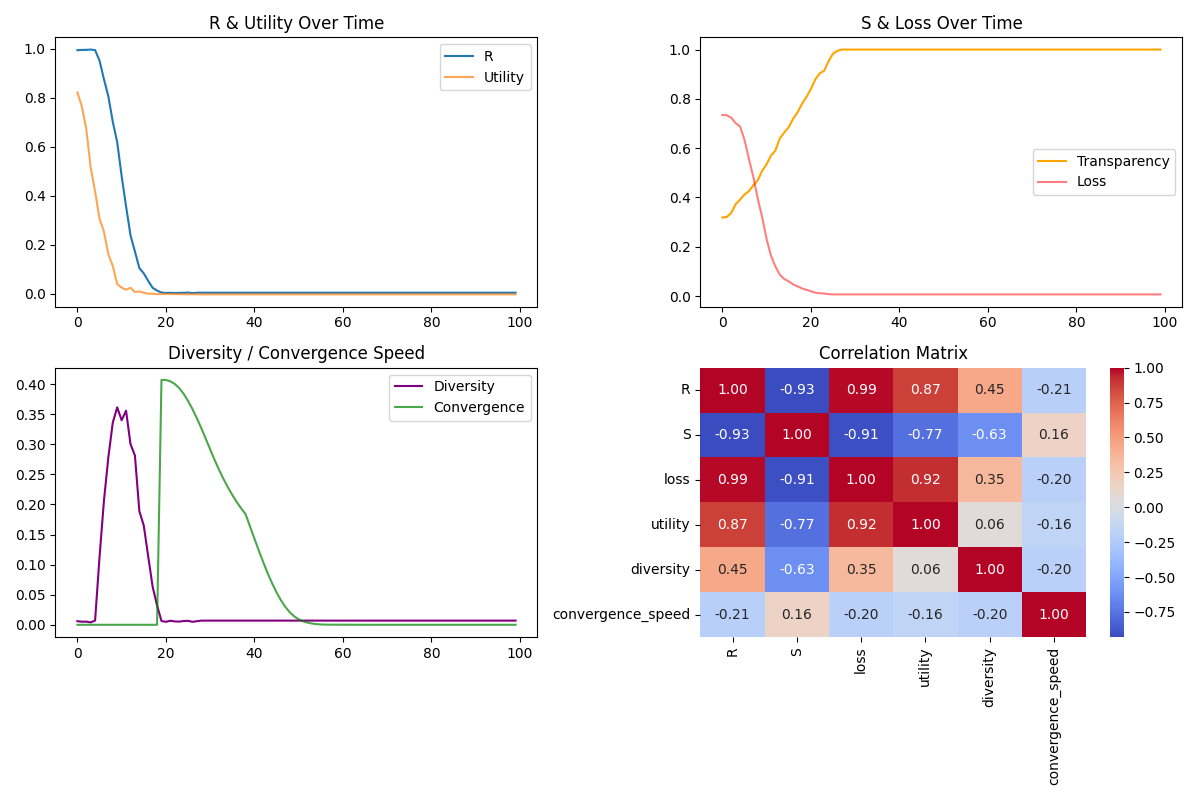 }
        \caption{General Experimental Results Part 2.}
        \label{fig:General Experimental Results Part 2}
    \end{subfigure}
    
    \caption{General Experimental Results.}
    \label{fig:General Experimental Results}
\end{figure}

\paragraph{1. Rent-Seeking and Utility Over Time.}
\begin{itemize}
    \item In the “R \& Utility Over Time” plot, the rent-seeking level \(R\) starts near 1.0 and then quickly drops to close to 0 between time steps 15 and 20, reflecting the strong suppressive effect of regulatory and reputation mechanisms on high-fake-content strategies.
    \item The average utility of agents initially remains relatively low, then stabilizes at a moderate level after the system reaches equilibrium. This indicates that the short-term gains from high rent-seeking are not sustainable in the long run.
\end{itemize}

\paragraph{2. Corresponding Changes in Platform Transparency and Social Loss.}
\begin{itemize}
    \item The “S \& Loss Over Time” plot shows that platform transparency \(S\) rises sharply from around 0.3 to nearly 1.0, while social loss decreases from a relatively high level to near 0.
    \item These trends align with the drop in \(R\), suggesting that under strong regulation and reputation constraints, the injection of fake information is substantially reduced, thereby lowering overall social costs.
\end{itemize}

\paragraph{3. Correlation of Multi-Dimensional Metrics and Diversity/Convergence.}
\begin{itemize}
    \item The correlation matrix reveals significant negative or positive relationships among rent-seeking, platform transparency, and social loss, consistent with theoretical expectations.
    \item The “Diversity / Convergence Speed” chart indicates that agents display high strategic diversity in the early phase, which rapidly declines around step 20. The system then maintains low diversity, implying that most agents converge to a stable low-rent-seeking strategy.
\end{itemize}

\paragraph{4. Evolution of Agent Investment and Reputation Distribution.}
\begin{itemize}
    \item Comparing the “R Dist” and “Rep Dist” figures shows that agents initially adopt high levels of fake-content injection, with reputation concentrated at a high level. Around \(t=50\), some agents reduce their injection due to detection and penalties, causing reputation to diverge.
    \item By \(t=99\), overall injection levels converge to a lower state, yet the reputation distribution remains somewhat polarized. This suggests that under strong regulation and reputation constraints, distinct groups of agents emerge based on their long-term behavior.
\end{itemize}

\begin{figure}[h]
  \includegraphics[width=\columnwidth]{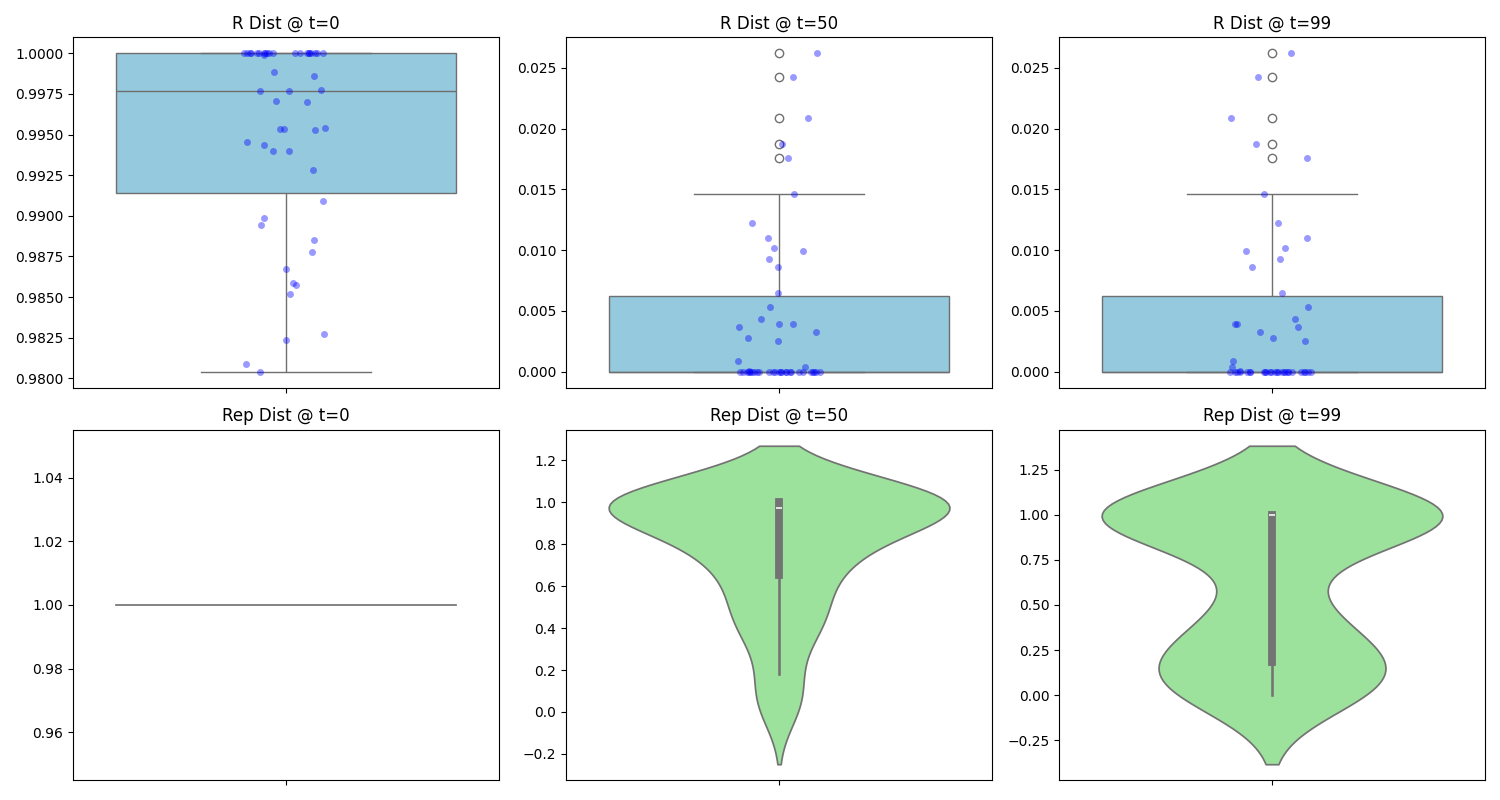}
  \caption{Evolution of Agent Investment and Reputation Distribution.}
  \label{fig:Evolution of Agent Investment and Reputation Distribution.}
\end{figure}

\paragraph{5. Phase Space and Temporal Cross-Section Analysis.}
\begin{itemize}
    \item The phase space plot (“Phase Space, S vs R”) visualizes the relationship between platform transparency and rent-seeking, with color denoting average utility. As \(S\) increases and \(R\) decreases, utility changes accordingly. This indicates that in the region of high transparency and low rent-seeking, agents do not achieve substantial long-term gains, while short-term benefits in the high-rent-seeking region are difficult to sustain.
    \item The “Highlight Major R Drops” plot underscores a significant decrease in \(R\) occurring between time steps 15 and 20, corresponding to the primary effect of regulation or reputation penalties within that period.
\end{itemize}

\begin{figure}[ht]
  \includegraphics[width=\columnwidth]{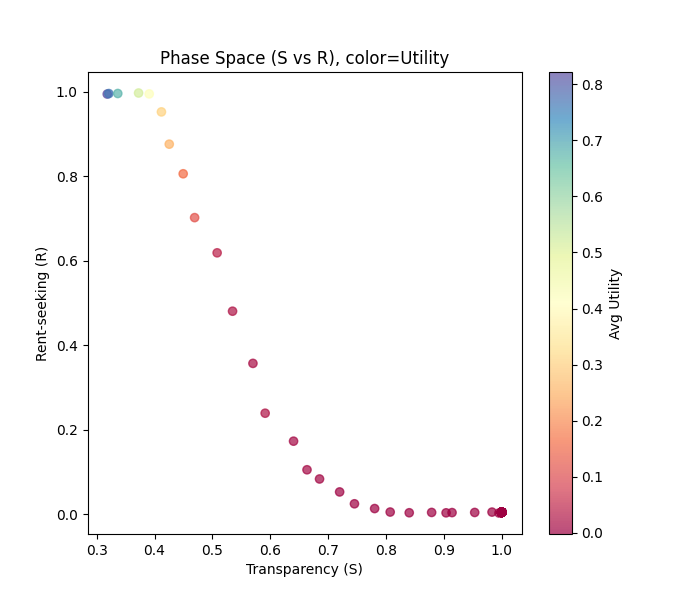}
  \caption{Phase Space and Temporal Cross-Section Analysis}
  \label{fig:Phase Space and Temporal Cross-Section Analysis}
\end{figure}

\begin{figure}[ht]
  \includegraphics[width=\columnwidth]{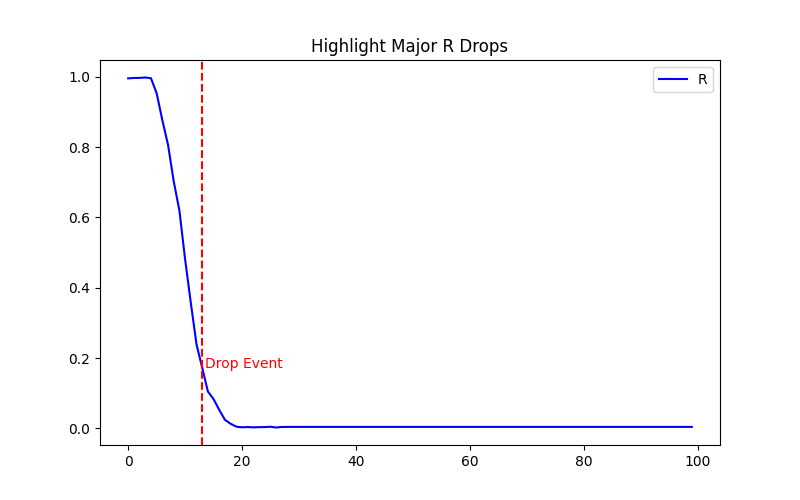}
  \caption{Highlight Major R Drops.}
  \label{fig:General experimental results.}
\end{figure}

Overall, these results confirm that under the current model assumptions and parameter settings, moderate and effective regulatory and reputation mechanisms can significantly curb fake information injection within a short period, raise platform transparency, and reduce social loss. Although agent behavior converges to a “low-rent-seeking, high-transparency” steady state, disparities in reputation still exist among different agents. These findings offer valuable insights for more complex scenario simulations and real-world platform governance, indicating that while reinforcing supervision and enhancing transparency are crucial, it is also important to balance potential group polarization and the risk of misclassification.

\subsection{Sensitive Analysis}

\begin{figure}[htbp]
    \centering
    
    \begin{subfigure}[b]{\columnwidth}
        \centering        
        \includegraphics[width=\textwidth]{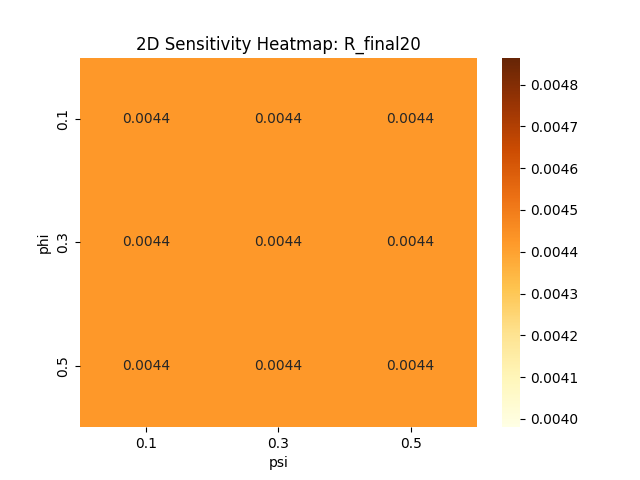}
        \caption{2D Sensitivity Heatmap: R.}
        \label{fig:2D Sensitivity Heatmap: R.}
    \end{subfigure}
    
    \begin{subfigure}[b]{\columnwidth}
        \centering
        \includegraphics[width=\textwidth]{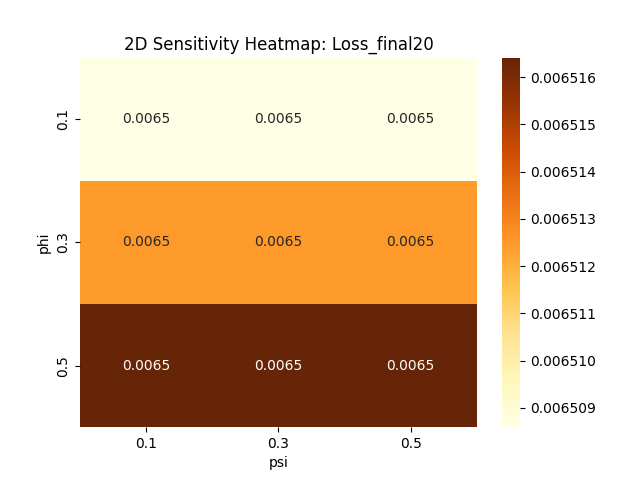}
        \caption{2D Sensitivity Heatmap: Loss.}
        \label{fig:2D Sensitivity Heatmap: Loss.}
    \end{subfigure}   
    
    \caption{Sensitivity Analysis Results.}
    \label{fig:Sensitivity Analysis Results.}
\end{figure}

From the two 2D sensitivity heatmaps, we can observe that when varying the values of \(\phi\) (the rent-seeking loss weight) and \(\psi\) (the transparency loss weight), there is no significant difference in either the final rent-seeking level (e.g., \(R\_\text{final20}\) at \(t=20\)) or the final social loss (\(\text{Loss}\_\text{final20}\)). In the first heatmap, \(R\_\text{final20}\) remains virtually the same across all parameter combinations, while in the second heatmap, \(\text{Loss}\_\text{final20}\) only fluctuates within a very narrow range.

This “uniform color block” phenomenon indicates that the system is not particularly sensitive to changes in \(\phi\) and \(\psi\). Even if the relative weights of rent-seeking losses or transparency losses are adjusted, both the final rent-seeking level and social loss remain at roughly the same low values. This suggests that once regulation intensity and reputation mechanisms are sufficiently effective, moderate variations in \(\phi\) and \(\psi\) do not substantially alter the system’s equilibrium state.

Overall, these heatmaps demonstrate a high degree of robustness with respect to \(\phi\) and \(\psi\). In the “high-transparency, low-rent-seeking” steady state, both rent-seeking behavior and social loss remain at low levels, largely unaffected by moderate changes in these parameters. From a policy-design perspective, this implies that as long as certain thresholds of regulatory strength and reputation penalties are maintained, the precise calibration of the “rent-seeking loss weight” and “transparency loss weight” does not need to be excessively fine-tuned. Under these conditions, effective control of fake content and a significant reduction in social loss can still be achieved.

\subsection{Conclusion}

In summary, our experiments demonstrate that rent-seeking behavior can be rapidly suppressed through moderate yet effective regulatory and reputation mechanisms. Within a short time, the system converges to a high-transparency, low-rent-seeking state, significantly reducing social losses. Although most agents adopt low-fake-content strategies, disparities in reputation levels remain, indicating a potential for group polarization.

\section{Discussion, Future Work, and Conclusion}

This study presents a dynamic economic model to analyze how generative AI transforms information rent-seeking behaviors and affects social welfare. We demonstrate that while generative AI can reduce traditional information rents by increasing transparency, it simultaneously introduces new rent-seeking avenues, such as algorithmic interference and large-scale information manipulation. Our analysis elucidates the equilibrium conditions governing the information environment, shows the delicate balance between transparency gains and new manipulation opportunities, and highlights the necessity of policy interventions to ensure positive outcomes.

Moving forward, several extensions can enrich the model. Incorporating bounded rationality, cognitive biases, or other behavioral factors can better capture real-world decision-making in complex information settings. Examining oligopolistic or monopoly conditions may yield different equilibrium outcomes, as market power interacts with generative AI-driven strategies. Furthermore, exploring how different AI modalities—such as vision, speech, and multimodal systems—jointly influence information rents and policy responses can provide a more comprehensive understanding of future challenges. Such extensions would make the model more applicable to diverse economic contexts and improve its policy relevance.

As generative AI capabilities advance, the information environment will continue to evolve, presenting new challenges and opportunities for economic actors. Continuous monitoring, timely policy updates, and active stakeholder engagement are crucial. By persistently refining our understanding and interventions, we can steer the information ecosystem toward greater fairness, efficiency, and societal well-being in the era of transformative AI technologies.

\section{Limitations}

While our model and analysis provide insightful theoretical conclusions, several limitations warrant attention.We employ reduced-form models for both agent behavior and generative AI effects. In reality, strategic interactions and agent heterogeneity may be more complex. The absence of large-scale, empirical datasets on generative AI-driven rent-seeking constrains our ability to calibrate parameters and validate model predictions against real-world scenarios. Acknowleding these limitations is the first step towards refining the framework and improving the robustness of future research.

\section{Acknowledgements}
During the writing of this article, generative artificial intelligence tools were used to assist in language polishing and literature retrieval. The AI tool helped optimize the grammatical structure and expression fluency of limited paragraphs, and assisted in screening research literature in related fields. All AI-polished text content has been strictly reviewed by the author to ensure that it complies with academic standards and is accompanied by accurate citations. The core research ideas, method design and conclusion derivation of this article were independently completed by the author, and the AI tool did not participate in the proposal of any innovative research ideas or the creation of substantive content. The author is fully responsible for the academic rigor, data authenticity and citation integrity of the full text, and hereby declares that the generative AI tool is not a co-author of this study.

\bibliographystyle{ACM-Reference-Format}
\bibliography{sample-base}

\appendix
\label{sec:appendix}

\section{Appendix: Theoretical Proofs}

In this appendix, we provide detailed proofs of the theoretical results discussed in the main text. These proofs establish the existence, uniqueness, and stability of the steady-state equilibrium and provide insights into the impact of policy interventions on rent-seeking behavior. We also discuss the limitations of the model assumptions and suggest possible extensions to improve its applicability.

\subsection{A.1 Existence and Uniqueness of the Steady-State Rent-Seeking Level}
\label{appendix:existence_uniqueness}

Having defined the basic model setup, we now examine the existence and uniqueness of the steady-state equilibrium. 

\paragraph{Proposition:} Under the assumptions of monotonic and concave benefit functions, strictly convex cost functions, and parameter conditions \( \gamma \geq 0 \) and \( \delta > 0 \), there exists a unique steady-state equilibrium \( (R^*, S^*) \) that satisfies the conditions:
\[
R^* = \frac{\gamma}{\delta}, \quad S_{t+1} = S_t = S^*.
\]

\paragraph{Proof:}  
To analyze the steady state, we begin by considering the state transition equation for the information environment:
\[
S_{t+1} = S_t + \gamma - \delta \overline{R_t} + \varepsilon_t.
\]
At steady state, we assume \( \mathbb{E}[\varepsilon_t] = 0 \) and \( S_{t+1} = S_t = S^* \), leading to the equation:
\[
0 = \gamma - \delta R^* \implies R^* = \frac{\gamma}{\delta}.
\]

The existence of \( R^* \) is guaranteed by the conditions \( \gamma \geq 0 \) and \( \delta > 0 \), ensuring that the ratio \( \frac{\gamma}{\delta} \) is well-defined and non-negative. Given the agents' optimization problem and the symmetry condition (where all agents choose the same rent-seeking effort \( R_{i,t}^* = R^* \) in equilibrium), substituting \( R^* \) into the first-order conditions confirms the existence of a unique \( R^* \). This ensures the consistency of the system.

Uniqueness follows from the strict convexity of the cost function and the monotonicity of the benefit function. If two distinct solutions \( R_1^* \) and \( R_2^* \) existed, they would both need to satisfy \( R^* = \frac{\gamma}{\delta} \), which contradicts the assumption of a unique solution. Therefore, the steady-state solution is unique.

Finally, \( S^* \) can be determined by substituting \( R^* \) back into the agent's equilibrium conditions and the steady-state information environment equation. The existence and uniqueness of \( S^* \) follow from the continuity and monotonicity properties of the state transition function and the well-defined equilibrium choice \( R^* \).

\textbf{Q.E.D.}

\subsection*{A.2 Comparative Statics on Policy Parameters}
\label{appendix:comparative_statics}

Next, we explore the impact of a per-unit tax \( \tau \) on rent-seeking behavior.

\paragraph{Proposition:} Suppose a per-unit tax \( \tau \) is introduced on rent-seeking. Under regularity conditions (differentiability, strict concavity of the net benefit function), an increase in \( \tau \) will reduce the equilibrium rent-seeking level \( R^* \).

\paragraph{Proof:}  
Let \( B_{\text{AI}}(R, S) \) denote the net benefit from rent-seeking, and \( C(R) \) represent the cost function. With a tax \( \tau R \), the first-order condition (FOC) at equilibrium becomes:
\[
\frac{\partial B_{\text{AI}}(R, S)}{\partial R} - \frac{\partial C(R)}{\partial R} - \tau = 0.
\]

As \( \tau \) increases, the marginal net benefit of rent-seeking decreases, causing agents to reduce their rent-seeking effort \( R^* \) to maintain optimality. Since \( \frac{\partial^2 C}{\partial R^2} > 0 \), the equilibrium rent-seeking level \( R^* \) must decrease as \( \tau \) increases. By the implicit function theorem, we have \( \frac{dR^*}{d\tau} < 0 \), confirming that the tax reduces the equilibrium rent-seeking level.

\textbf{Q.E.D.}

\subsection*{A.3 Stability of the Steady State}
\label{appendix:stability}

We now examine the stability of the steady-state equilibrium \( (R^*, S^*) \) under mild conditions on the parameters.

\paragraph{Proposition:} The steady-state equilibrium \( (R^*, S^*) \) is locally stable under reasonable assumptions on the parameters.

\paragraph{Proof (Sketch):}  
Let \( \hat{S}_t = S_t - S^* \) and \( \hat{R}_t = R_t - R^* \) represent the deviations from the steady state. Linearizing the state transition equation around \( (R^*, S^*) \):
\[
\hat{S}_{t+1} \approx \hat{S}_t - \delta \hat{R}_t,
\]
since \( \gamma \) and \( S^* \) are constants at equilibrium, and \( \varepsilon_t \) has mean zero.

Similarly, linearizing the agent's first-order condition (FOC) around \( R^* \) and \( S^* \), we get:
\[
\hat{R}_t \approx -\frac{f_S}{f_R} \hat{S}_t,
\]
where \( f_S \) and \( f_R \) are partial derivatives of the net marginal benefit function with respect to \( S \) and \( R \), evaluated at \( (R^*, S^*) \). Under typical conditions (e.g., \( f_R < 0 \), ensuring a unique maximum for the agent's problem), the system is stable. The linear system of equations for \( \hat{R}_t \) and \( \hat{S}_t \) will produce eigenvalues with magnitudes less than one, ensuring that small deviations from equilibrium dissipate over time and the system returns to equilibrium.

\textbf{Q.E.D.}

\subsection*{A.4 Policy Optimality Conditions}
\label{appendix:policy_optimality}

Finally, we consider the regulator's problem of minimizing social welfare losses by choosing optimal policy interventions.

\paragraph{Proposition:} The optimal tax rate \( \tau_t^* \) that minimizes social welfare losses satisfies:
\[
\frac{\partial}{\partial \tau_t} \mathbb{E}\left[\sum_{t=0}^{T} \delta^t L(R_t, S_t)\right] = 0,
\]
where \( L(R_t, S_t) = \phi R_t^2 + \psi (1 - S_t)^2 \) is the social welfare loss function.

\paragraph{Proof (Sketch):}  
Applying the chain rule, we obtain:
\[
\frac{\partial L}{\partial \tau_t} = \frac{\partial L}{\partial R_t}\frac{\partial R_t}{\partial \tau_t} + \frac{\partial L}{\partial S_t}\frac{\partial S_t}{\partial \tau_t}.
\]

At equilibrium, \( R_t \) and \( S_t \) are functions of \( \tau_t \). As \( \tau_t \) increases, the marginal rent-seeking effort \( R_t \) decreases, leading to lower opacity and resource misallocation. The optimal tax \( \tau_t^* \) balances the reduction in rent-seeking with the potential negative effects on information transparency. Numerical methods or sensitivity analyses can be employed to identify the optimal \( \tau_t^* \) that minimizes welfare losses.

\subsection{Discussion}
These theoretical results form the foundation for the numerical simulations and policy analysis presented in the main text. They establish the equilibrium conditions, show the effects of policy interventions on rent-seeking behavior, and highlight the importance of balancing transparency with rent-seeking control. These insights will guide further simulations and help refine policy recommendations.

\section{Detailed Experimental Setup and Parameters}

This appendix provides a comprehensive description of the experimental framework, including detailed parameters and simulation design. The framework is built on the background of “Generative AI + Rent-Seeking Behavior” and focuses on the scenario of fake information and intelligence manipulation on social media platforms. It captures the dual effects of generative AI in both content generation and moderation, while reflecting the dynamic game among information asymmetry, agent rent-seeking, and social welfare losses. The goal is to explore the optimal balance among various policy tools (such as taxation, detection, and disclosure) under multi-round and multi-scenario settings.

\subsection{Experiment Scenario and Overall Objectives}

\paragraph{Scenario Overview.} 
The simulation models a social media environment where fake information and intelligence manipulation occur. The primary participants include:
\begin{itemize}
    \item Ordinary users and content producers, who browse and post content to obtain accurate information or entertainment.
    \item AI agents (rent-seekers) that employ generative AI to mass-produce content (which may be true or fake) in order to secure economic or political benefits (for example, by influencing public opinion to affect stock prices, brand reputation, or political outcomes).
    \item The platform or regulator, which monitors and governs content through measures such as content review, algorithmic detection, and policy interventions to enhance overall information transparency and reduce societal harm.
\end{itemize}

\paragraph{Experimental Objectives.}
The experiment aims to:
\begin{enumerate}
    \item Enhance realism by extending the traditional scenario of fake information and intelligence manipulation. Long-term agent payoffs, reputation penalties, and potential "collateral damage costs" due to stringent regulation are incorporated to produce more realistic simulation results.
    \item Explore policy tools by running multi-round and multi-scenario numerical simulations. The goal is to identify the optimal combination and boundary conditions for various policy instruments (e.g., taxation, detection probability, and information disclosure intensity) under a controlled environment with 50 to 100 agents.
    \item Reveal the game dynamics by simulating the dual role of generative AI in information generation, propagation, and moderation. The experiment examines how agents balance immediate gains with future risks (such as reputation deterioration and fines), thereby uncovering the trade-off between new AI-enabled rent-seeking behaviors and social welfare losses.
\end{enumerate}

\subsection{Model Structure and Key Modifications}

\paragraph{Multi-Agent Structure.} 
The model assumes a set of $N$ agents (typically between 50 and 100). Each agent selects a fake information injection amount $R_{i,t}$ in each round to maximize its illicit gains. A subset of these agents, such as traditional trolls, may have lower gains but still impact the information environment. Ordinary users and content producers are treated as an exogenous variable that indirectly influences the collateral damage cost. The platform or regulator implements interventions, such as taxation, detection, and information disclosure, to maintain overall information transparency denoted by $S_t$.

\paragraph{Long-Term Decision-Making and Discounting.} 
Each agent’s long-term objective is to maximize its cumulative discounted payoff:
\[
U_i = \sum_{t=0}^{T} \beta^t \cdot \pi_{i,t},
\]
where the discount factor $\beta \in (0,1)$ reflects the agent’s valuation of future returns, necessitating a trade-off between immediate gains and long-term risks (e.g., reputation loss or fines).

\paragraph{Reputation Mechanism.} 
Each agent maintains a reputation score, $\text{rep}_{i,t} \in [0,1]$, which is initially set to 1. If an agent is detected generating fake content in a given round, its reputation is updated as
\[
\text{rep}_{i,t+1} = \max\bigl(\text{rep}_{i,t} - \Delta_{\text{rep}},\, 0\bigr),
\]
where $\Delta_{\text{rep}}$ (e.g., between 0.1 and 0.2) is the reputation deduction. A lower reputation reduces the efficiency of fake information propagation (by, for instance, multiplying the revenue function by $\text{rep}_{i,t}$) and increases the likelihood of future detection.

\paragraph{Platform Information Transparency and Generative AI Effects.} 
The platform’s transparency is represented by $S_t \in [0,1]$, indicating the proportion of true and verifiable information. Higher transparency makes it easier for users to distinguish real from fake content, thereby hindering the spread of misinformation. Two parameters capture the dual role of generative AI: $\gamma$ denotes the transparency gain from automated review and content traceability, while $\delta$ reflects the platform’s susceptibility to fake content—indicating how quickly AI-generated misinformation can erode transparency.

\paragraph{Policy Side Effects.} 
Additional costs are incorporated in both agent payoffs and social welfare losses to reflect collateral damage, such as the misclassification of true content or the suppression of innovation caused by strict regulation. For example, the social welfare loss function includes a term:
\begin{align*}
L_{\text{total}}(t) &= \phi \left(\frac{1}{N} \sum_{i=1}^{N} R_{i,t}\right)^2 \\
&\quad + \psi (1 - S_t)^2 \\
&\quad + \eta \cdot \text{CollateralDamage}_t.
\end{align*}
where $(\text{Collateral Damage})_t$ is estimated based on the platform’s detection accuracy.

\subsection{Core Equations of the Modified Model}

\paragraph{Agent Payoff Function.} 
The payoff for each agent is defined as:
\begin{align*}
\pi_{i,t} &= B_{\text{AI}}\left(R_{i,t}, S_t, \text{rep}_{i,t}\right) 
- C\left(R_{i,t}\right) \\
&\quad - \tau \cdot R_{i,t} - \text{Fines}_{i,t}.
\end{align*}

Here, 
\begin{align*}
B_{\text{AI}}\bigl(R_{i,t}, S_t, \text{rep}_{i,t}\bigr) &= 
\Bigl[\beta_1 R_{i,t}(1 - S_t) \\
&\quad + \Delta B_{\text{AI}}(R_{i,t}, S_t)\Bigr] 
\times \text{rep}_{i,t}.
\end{align*}

represents the additional gains from generative AI (with reputation incorporated), $C(R_{i,t})$ is the cost associated with producing fake content, $\tau \cdot R_{i,t}$ is the tax imposed, and $\text{Fines}_{i,t}$ is determined based on the detection probability $\theta$ and a penalty multiplier $\kappa$. Agents aim to maximize their cumulative discounted payoff.

\paragraph{Platform Transparency Update Equation.} 
The update rule for the platform’s transparency is given by:
\[
S_{t+1} = S_t + \gamma - \delta \cdot \overline{R_t} + \varepsilon_t,
\]
where $\overline{R_t}$ is the average fake content injection by all agents in round $t$, $\gamma$ is the transparency gain from fact-checking or automated review, $\delta$ reflects the susceptibility to misinformation, and $\varepsilon_t$ is a noise term representing external disturbances.

\paragraph{Social Welfare Loss Function.} 
The overall social welfare loss is defined as:

\begin{align*}
L_{\text{total}}(t) &= \phi \left(\frac{1}{N} \sum_{i=1}^{N} R_{i,t}\right)^2 
+ \psi (1 - S_t)^2 \\
&\quad + \eta \cdot \text{CollateralDamage}_t.
\end{align*}

The first term captures the impact of misinformation on resource allocation and user attention, the second term represents the loss of social trust and information quality, and the third term accounts for additional societal costs due to misclassification of true content under strict regulation.

\subsection{Experimental Design and Simulation Procedure}

\paragraph{Environment Initialization.} 
The simulation is set to run for $T=100$ discrete time steps, modeling the platform’s evolution over 100 rounds. The number of agents is initially set to $N=100$, although a mix of traditional trolls can be included. The initial platform transparency is set to $S_0=0.5$, indicating a mix of true and false content. Key parameters such as $\gamma$, $\delta$, the discount factor $\beta$, reputation deduction $\Delta_{\text{rep}}$, base detection probability $\theta_0$, and penalty multiplier $\kappa$ are preliminarily determined based on pilot studies.

\paragraph{Agent Decision Process.} 
In each round, every agent decides its fake content injection $R_{i,t}$ based on the current platform transparency $S_t$, its reputation $\text{rep}_{i,t}$, and other regulatory parameters. Agents compute their payoff using

\begin{align*}
\pi_{i,t} &= B_{\text{AI}}\bigl(R_{i,t}, S_t, \text{rep}_{i,t}\bigr) 
- C\bigl(R_{i,t}\bigr) \\
&\quad - \tau \cdot R_{i,t} - \text{Fines}_{i,t}.
\end{align*}

and factor in future returns using the discount factor $\beta$. Two decision algorithms are considered: one based on reinforcement learning (e.g., Q-learning or policy gradient methods) where agents learn optimal actions based on the state $(S_t, \text{rep}_{i,t}, \tau, \theta)$, and another based on piecewise optimization via local iterative search.

\paragraph{Reputation and Detection Module.} 
If an agent is detected producing fake content in round $t$, its reputation is updated according to
\[
\text{rep}_{i,t+1} = \max\bigl(\text{rep}_{i,t} - \Delta_{\text{rep}},\, 0\bigr).
\]
If no detection occurs, reputation remains constant or recovers slowly. The detection probability for an agent is defined as
\[
\theta_{i,t} = \theta_0 + \alpha \cdot (1 - \text{rep}_{i,t}),
\]
implying that lower reputation increases the chance of detection.

\paragraph{Platform and Regulatory Interventions.} 
The platform imposes taxation and fines on agents detected with fake content; fines can be computed as either $\tau \cdot R_{i,t}$ or $\kappa \cdot R_{i,t}$. Information disclosure and review measures are applied to reduce the additional benefit $\Delta B_{\text{AI}}$, thereby improving transparency $S_t$. The platform’s transparency is dynamically updated based on the average fake content $\overline{R_t}$ and the external noise term $\varepsilon_t$.

\paragraph{Experimental Scenarios and Parameter Settings.} 
Several experimental scenarios are considered:
\begin{itemize}
    \item A baseline group without discounting, reputation mechanisms, or regulatory side effects.
    \item A long-term perspective group that incorporates discounting and reputation mechanisms to observe long-term behavioral changes.
    \item Groups under strong versus weak regulation, where high-tax/high-detection settings are compared with low-tax/low-detection settings.
    \item A collateral damage comparison group where the misclassification rate $\rho$ (e.g., 0.01, 0.05, 0.1) is varied to assess its impact on social welfare.
\end{itemize}
Key parameters are explored via grid or random search:
\[
\beta \in \{0.8, 0.9, 0.95\}, \quad \Delta_{\text{rep}} \in \{0.1, 0.2\}, \quad 
\]
\[
\rho \in \{0.01, 0.05, 0.1\}, \quad \tau \in \{0, 0.1, 0.3\}.
\]
\[
\theta_0 \in \{0.2, 0.5, 0.8\}, \quad \kappa \in \{2, 5\}.
\]

\paragraph{Output Metrics.} 
The main output indicators include:
\begin{itemize}
    \item The average fake information injection $\overline{R_t}$ over time, to determine whether the system reaches a steady state or exhibits cyclic behavior.
    \item The evolution of platform transparency $S_t$, which reflects the overall quality of information on the platform.
    \item The distribution of agent reputations $\text{rep}_{i,t}$ and its relation to individual payoffs.
    \item The instantaneous and cumulative social welfare loss $L_{\text{total}}(t)$ under different regulatory policies.
    \item The collateral damage rate, defined as the proportion of true content mistakenly affected by stringent regulation.
\end{itemize}

\subsection{Summary}

The experimental framework presented here constructs a multi-agent simulation environment that integrates the dual role of generative AI in fake information generation and manipulation with agents' long-term payoffs, reputation mechanisms, and platform regulatory interventions (including detection, and information disclosure). By incorporating long-term decision-making, reputation penalties, dynamic transparency updates, and policy side effects such as collateral damage, the framework not only validates theoretical equilibrium relationships and rent allocation mechanisms but also provides quantitative insights for developing effective regulatory policies.

\end{document}